\documentclass{optica-article}

\journal{opticajournal} 

\articletype{Research Article}

\usepackage{lineno}
\linenumbers 

\usepackage{bm}

\usepackage{gensymb}
\usepackage{here}
\usepackage{svg}
\usepackage{soul}
\setstcolor{red}

\begin{document}
\nolinenumbers
\title{Phase-edge imaging using q-plate shifts for faster and simpler microscopy}

\author{Jigme Zangpo$^1$, Hirokazu Kobayashi$^2$, and Ryo Yasuhara$^{1,}$$^{3,}$$^*$}

\address{$^1$National Institute for Fusion Science,
322-6 Oroshi-cho, Toki City, Gifu, Japan\\
$^3$Graduate School of Engineering, Kochi University of Technology, 185 Miyanokuchi, Tosayamada, Kami City, Kochi 782-8502, Japan\\
$^2$The Graduate University for Advanced Studies, SOKENDAI, 322-6 Oroshi-cho, Toki, Gifu, Japan}

\email{\authormark{*}yasuhara.ryo@nifs.ac.jp} 


\begin{abstract*} 
We present a simplified method for isolating the edges of a phase object from the edges of an amplitude object using a $4f$ system with an off-axis q-plate. Instead of the four off-axis shifts of the q-plate required in previous work, we need only two shifts (along $\pm x$) combined with linear polarizers at $45^\circ$ and $135^\circ$. The number of measurements is reduced by half, potentially doubling the acquisition speed. We derive the theoretical basis, showing that the resulting intensity corresponds to the phase gradient squared, with amplitude‑object contributions eliminated. Experiments on two phase‑amplitude object samples demonstrate amplitude‑edge reduction up to $97.6\%$ and correlation coefficients up to $0.78$ (sample 1) and $0.75$ (sample 2). In overlapping regions, the phase edge is partially recovered; full recovery would require additional processing such as inverse filtering. This research is useful for biological imaging applications where fast and simple phase‑edge isolation is desired.

\end{abstract*}

\section{Introduction}
Microscopy is an important tool in biological and material sciences, which provides detailed structures at the micro- and nanoscale. Traditional bright-field microscopy displays intensity contrast; however, it provides low contrast for transparent specimens. Phase-contrast methods like Zernike phase contrast and differential interference contrast (DIC) can observe transparent specimens. However, Zernike phase contrast lacks a direct edge-enhancement technique, and DIC detects edges only in one direction \cite{zernike1942phase,DIC1}. Edge enhancement is widely used in fields such as optical information processing \cite{offaxis1}, biomedical imaging \cite{biological1}, the medical field \cite{medical}, astronomy \cite{astonomy1}, and single-pixel imaging \cite{VPM4phase1,zangpo2026single}, as highlighting contours improves feature visibility.

F\"urhapter \textit{et al.} introduced spiral phase contrast microscopy, which can achieve all-directional edge enhancement for both phase and amplitude objects \cite{2005spiral}. This method employs a $4f$ system with a vortex phase filter $e^{i l \phi}$ at the Fourier plane, where $l$ is the topological charge and $\phi=\arctan(y/x)$ is the azimuthal angle \cite{spiralPF4,spiralPF7,2024isolation}. Vortex filters are classified into two types: scalar (polarization-independent) and vectorial (polarization-dependent). For complex specimens that contain both phase and amplitude objects (phase-amplitude objects), vectorial vortex filtering using filter like q-plates enhances edges effectively while minimizing the interference between phase and amplitude contributions that is often present in scalar vortex filtering \cite{qplate2,zangpo2023edge,zangpoedge}. However, vectorial vortex filtering enhances the edges of both phase objects and amplitude objects equally, which can make them difficult to distinguish \cite{zangpo2023edge}. Reference~\cite{zangpo2022isolation,2024isolation} demonstrated that phase-object edges can be isolated by combining images from four distinct shifts of an off-axis q-plate. While this method is effective, its requirement for four separate measurements may limit imaging speed and increase the risk of errors due to multiple shifts of the q-plate.

In this paper, we introduce a method for isolating phase edges using only two shifts of an off-axis q-plate with $45^\circ$ and $135^\circ$ polarizers, which reduces the number of required measurements by half compared to the prior four-shift method, potentially doubling the acquisition speed. We present a theoretical analysis for this two-shift method and validate it experimentally on two phase-amplitude object samples. Our results show amplitude-edge reduction up to $97.6\%$ in the isolated phase edge image and a correlation coefficient up to $0.78$ between on-axis and off-axis phase edges. This work represents a step towards making phase-edge isolation more practical for biological imaging.

\section{Theoretical analysis}

\begin{figure}[htp!]
\centering
\includegraphics[width=\linewidth]{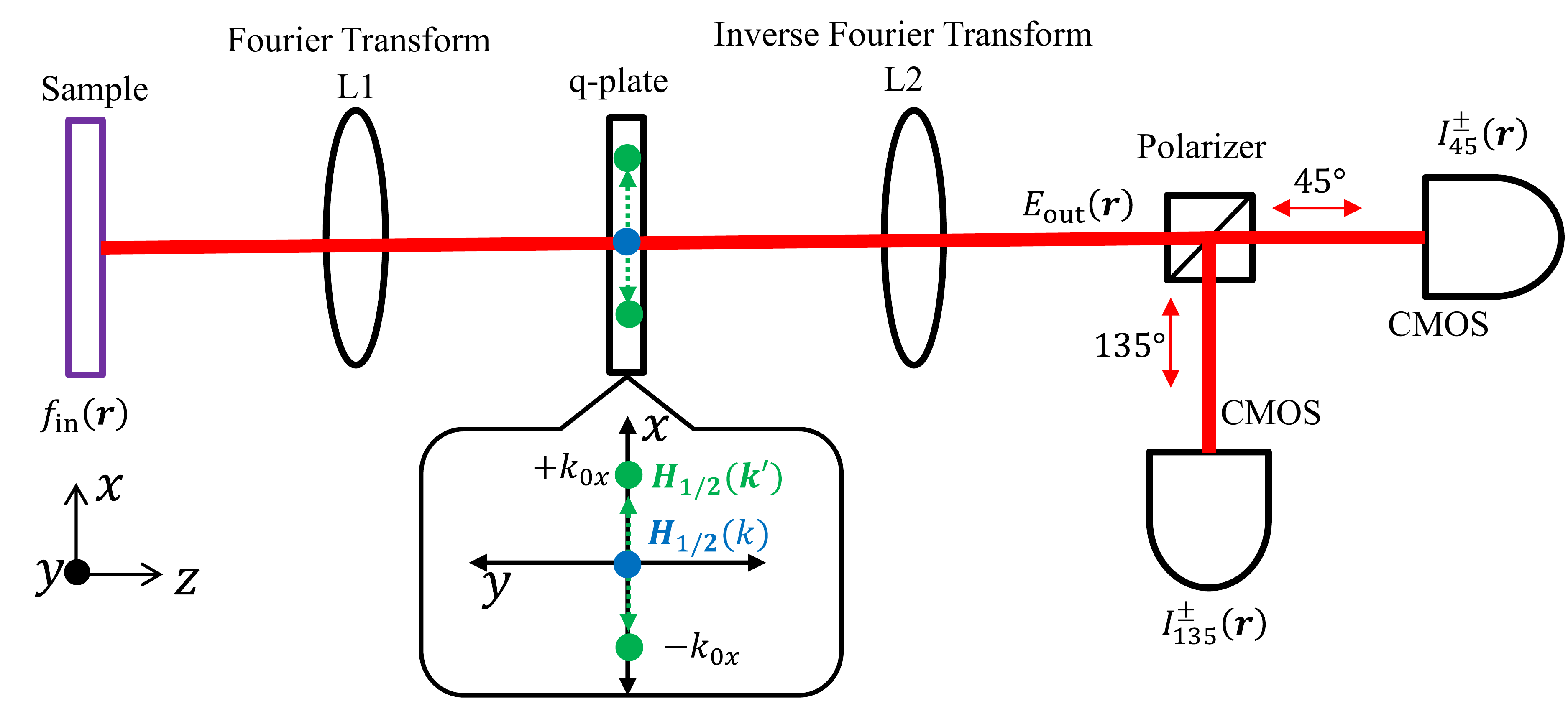}
\caption{$4f$ imaging system with the q-plate.}
\label{fig_4fsystem}
\end{figure}

The $4f$ system with a q-plate ($q=1/2$) is shown in Fig.~\ref{fig_4fsystem}. The input object $f_{\text{in}}(\bm{r}) = A(\bm{r}) e^{iB(\bm{r})}$ is illuminated by horizontally polarized light $\bm{E}_{\text{in}} = (1,0)^T$, where $\bm{r} = (x,y)$ denotes the transverse spatial coordinate, $A(\bm{r})$ is the amplitude object and $B(\bm{r})$ is the phase object. The first lens (L1) performs a Fourier transform; at the Fourier plane, the spectrum is multiplied by the q-plate transmission function; the second lens (L2) then performs an inverse Fourier transform. A linear polarizer is placed before the CMOS camera.

The transmission function of the q-plate in the Fourier domain is
\begin{equation}
\bm{H}_{1/2}(\bm{k}) = \frac{i}{k_r} \begin{pmatrix} k_x & k_y \\ k_y & -k_x \end{pmatrix}, \quad k_r = \sqrt{k_x^2 + k_y^2},
\label{Eq1}
\end{equation}
where $\bm{k} = (k_x, k_y)$ are the spatial angular frequencies (i.e., coordinates in the Fourier plane).

To isolate phase‑object edges, we displace the q-plate by $k_0$ along the $x$-direction. The shifted transmission function becomes
\begin{equation}
\bm{H}_{1/2}(\bm{k}') = \frac{i}{k_r'} \begin{pmatrix} k_x' & k_y \\ k_y & -k_x' \end{pmatrix},
\label{Eq2}
\end{equation}
with $k_x' = k_x \pm k_0$ and $k_r' = \sqrt{k_x'^2 + k_y^2}$ (the $\pm$ denotes positive or negative shift).

The object spectrum after L1 is $F_{\text{in}}(\bm{k})\bm{E}_{\text{in}}$, where $F_{\text{in}}(\bm{k})$ is the Fourier transform of $f_{\text{in}}(\bm{r})$. This spectrum is multiplied by the shifted filter $\bm{H}_{1/2}(\bm{k}')$ and then inverse‑transformed by L2. The shift introduces an extra phase $e^{\pm i k_0 x}$ in the image plane (see Ref.~\cite{2024isolation} for details). The resulting field incident on the polarizer is denoted $\mathbf{E}_{\text{out}}(\bm{r}, \pm k_0)$. Using the differential and frequency‑shifting property of the Fourier transform, and neglecting the $1/r$ convolution (ideal‑edge approximation), its Cartesian components for a shift $k_0$ along $x$ are given by
\begin{equation}
\label{E3}
\mathbf{E}_{\text{out}}^{\pm}(\bm{r}) = \bm{\nabla} f_{\text{in}}(\bm{r}) \mp i k_0 f_{\text{in}}(\bm{r}) \, \hat{\bm{e}}_x,
\end{equation}
where $\bm{\nabla} = (\partial_x, \partial_y)^T$ is the gradient operator, and $\hat{\bm{e}}_x = (1,0)^T$ is the unit vector along the $x$-direction. Here, $\mathbf{E}_{\text{out}}^{+}$ corresponds to a positive shift $+k_0$ and $\mathbf{E}_{\text{out}}^{-}$ to a negative shift $-k_0$.

For each displacement, a linear polarizer oriented at $45^\circ$ or $135^\circ$ projects the field, and the CMOS camera records the intensity. The measured intensities are given by the absolute squares of the projected fields:
\begin{equation}
\label{E4}
I_{45}^{\pm}(\bm{r}) = \frac{1}{2} \bigl| \mathbf{P}_{45}^T \cdot \mathbf{E}_{\text{out}}(\bm{r}, \pm k_0) \bigr|^2, \qquad
I_{135}^{\pm}(\bm{r}) = \frac{1}{2} \bigl| \mathbf{P}_{135}^T \cdot \mathbf{E}_{\text{out}}(\bm{r}, \pm k_0) \bigr|^2.
\end{equation}
Forming the differences $\Delta_{45} = I_{45}^{+} - I_{45}^{-}$ and $\Delta_{135} = I_{135}^{+} - I_{135}^{-}$ and combining them as $I_{\text{PO}}(\bm{r}) = (\Delta_{45})^2 + (\Delta_{135})^2$, we obtain the isolated phase edge:
\begin{equation}
I_{\text{PO}}(\bm{r}) = 8k_0^2 A(\bm{r})^4 \, |\bm{\nabla} B(\bm{r})|^2.
\label{Eq5}
\end{equation}

Thus, both gradient components $\frac{\partial B}{\partial x}$ and $\frac{\partial B}{\partial y}$ are recovered without the need for $y$-shifts, thanks to the mixing of field components by the $45^\circ$ and $135^\circ$ polarizers. In Eq.~\ref{Eq5}, the contributions from amplitude‑object edges are eliminated. The absolute squares in Eq.~\ref{E4} yield the recorded intensities, and the combination in Eq.~\ref{Eq5} produces the final isolated phase edge.

\section{Experimental setup and results}
\begin{figure}[htp!]
\centering
\includegraphics[width=\linewidth]{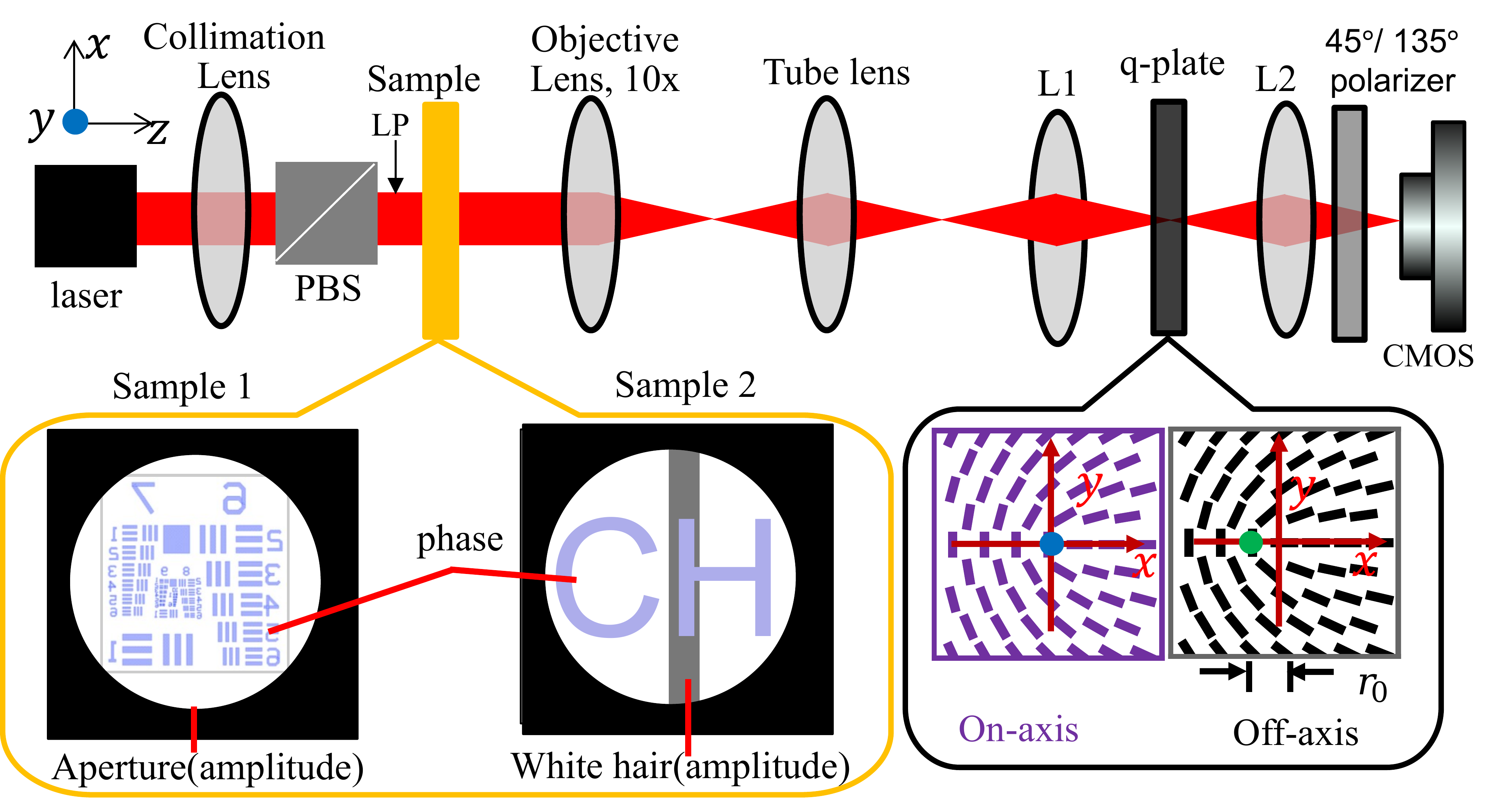}
\caption{Experimental setup and samples. PBS as polarizing beam splitter and LP denotes linearly polarized light.}
\label{exp_setup}
\end{figure}
The experimental configuration (Fig.~\ref{exp_setup}) follows the same bright-field microscope and $4f$ imaging architecture described in Ref.~\cite{2024isolation}, with two modifications: (i) only two off-axis displacements ($\pm x$) of the q-plate are used, and (ii) a rotating linear polarizer (mounted on a rotation mount) is sequentially set to $45^\circ$ or $135^\circ$ before the camera.

A 635~nm laser (Thorlabs HLS635) is collimated (Thorlabs AC254-050-A-ML) and then horizontally polarized by a polarizing beam splitter (Thorlabs CM1-PBS251). The bright-field microscope section consists of a $10\times$ objective (focal length $f_{\text{OB}}=18$~mm) and a tube lens ($f_{\text{TL}}=180$~mm). The second $4f$ system uses two Fourier lenses L1 and L2 with focal lengths $f_1=300$~mm and $f_2=250$~mm. A commercial q-plate (Thorlabs WPV10L-633, retardance $\pi$, topological charge $1$) is mounted on a translation stage that can shift the q-plate by $r_0$ along the $x$ direction. The displacement $r_0 = (f_1/k) k_0$ is controlled with a differential-drive $xy$ translator (precision $0.5$~$\mu$m per graduation). A linear polarizer (Thorlabs RSP1X15 rotation mount) is sequentially set to $45^\circ$ and $135^\circ$, and a CMOS camera (Thorlabs DCC1645C) records the final intensity.

Two phase‑amplitude object samples are prepared (see the sample images in Fig. \ref{exp_setup}). The first is a $350$~nm-thick USAF 1951 phase test target (refractive index 1.52) surrounded by a circular aperture (radius $300$~$\mu$m) that acts as the amplitude object. The phase and amplitude object do not overlap. The second sample adds a white hair (non-zero transmittance) that overlaps the letters “CH” of the phase test target; the aperture is also present. The hair is placed approximately $0.8$~mm out of focus relative to the phase test target.

\subsection{Results: Aperture surrounding phase test target}

\begin{figure}[htp!]
\centering
\includegraphics[width=\linewidth]{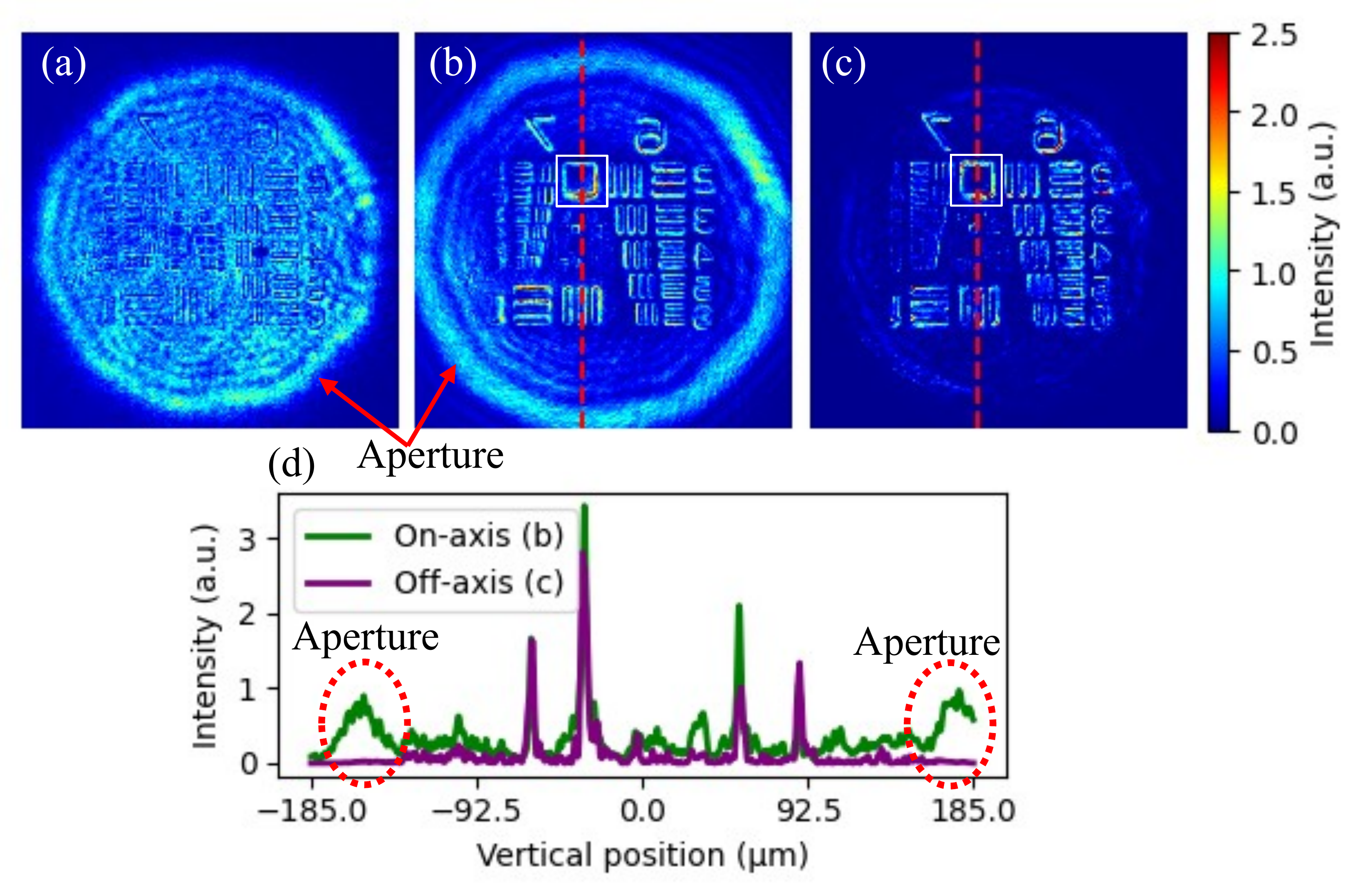}
\caption{Experimental results for the first phase-amplitude object (aperture surrounding phase test target). (a) Without q-plate, (b) on-axis edge enhancement, (c) isolated phase edge using the two-shift method, (d) vertical cross-section along the red dotted line.}
\label{result_aper}
\end{figure}

Figure~\ref{result_aper} shows the experimental results for the first phase‑amplitude object: a 350‑nm‑thick USAF 1951 phase test target surrounded by a circular aperture (amplitude object). Without a q-plate, the aperture is clearly visible but the phase test target remains nearly invisible (Fig.~\ref{result_aper}(a)). With an on‑axis q-plate, both the aperture edge and the phase edges are enhanced simultaneously (Fig.~\ref{result_aper}(b)). Using the proposed two‑shift method with $45^\circ$ and $135^\circ$ polarizers, the isolated phase edges are obtained as shown in Fig.~\ref{result_aper}(c). The amplitude edge is reduced, as illustrated in the vertical cross‑section plot along the red dotted in (b) and (c) in Fig.~\ref{result_aper}(d), where the green and purple curves correspond to the on‑axis (b) and off‑axis (c) images, respectively.

The reduction of the aperture edge was quantified by $R = \frac{I_{\text{on}} - I_{\text{off}}}{I_{\text{on}}}$, where $I_{\text{on}}$ is the average intensity of the aperture edge in the on‑axis image and $I_{\text{off}}$ is the corresponding intensity in the off‑axis isolated image. Table~\ref{tab:sample1} lists the edge reduction and the correlation coefficient (CC) between the isolated phase edges (off‑axis) and the phase edges extracted from the on‑axis image. The correlation coefficient was calculated for the square phase object enclosed by the white dotted square. Results are shown for five different displacements $r_0$ of the q-plate.

\begin{table}[htbp]
\centering
\caption{Edge reduction and correlation coefficient for sample 1 as a function of q-plate displacement.}
\label{tab:sample1}
\begin{tabular}{c|ccccc}
Displacement ($\mu$m) & 25 & 50 & 75 & 100 & 125 \\
\hline
Edge reduction (\%) & 94.46 & 96.66 & 96.98 & 97.36 & 97.58 \\
CC & 0.51 & 0.61 & 0.69 & 0.74 & 0.78 \\
\end{tabular}
\end{table}

The edge reduction increases with displacement, reaching $97.6\%$ at $r_0 = 125~\mu$m. The correlation coefficient also improves as the displacement increases, indicating better preservation of the phase edge structure at larger shifts. The correlation coefficient reaches 0.78 at 125~$\mu$m, which is lower than the 0.93 reported in the four-shift method~\cite{2024isolation}. This is most likely due to non-optimal overall alignment of the system, not a fundamental drawback of the two-shift technique. With improved alignment, we expect comparable performance.

\subsection{Results: Aperture surrounding phase test target and overlapping with white hair}

\begin{figure}[htp!]
\centering
\includegraphics[width=\linewidth]{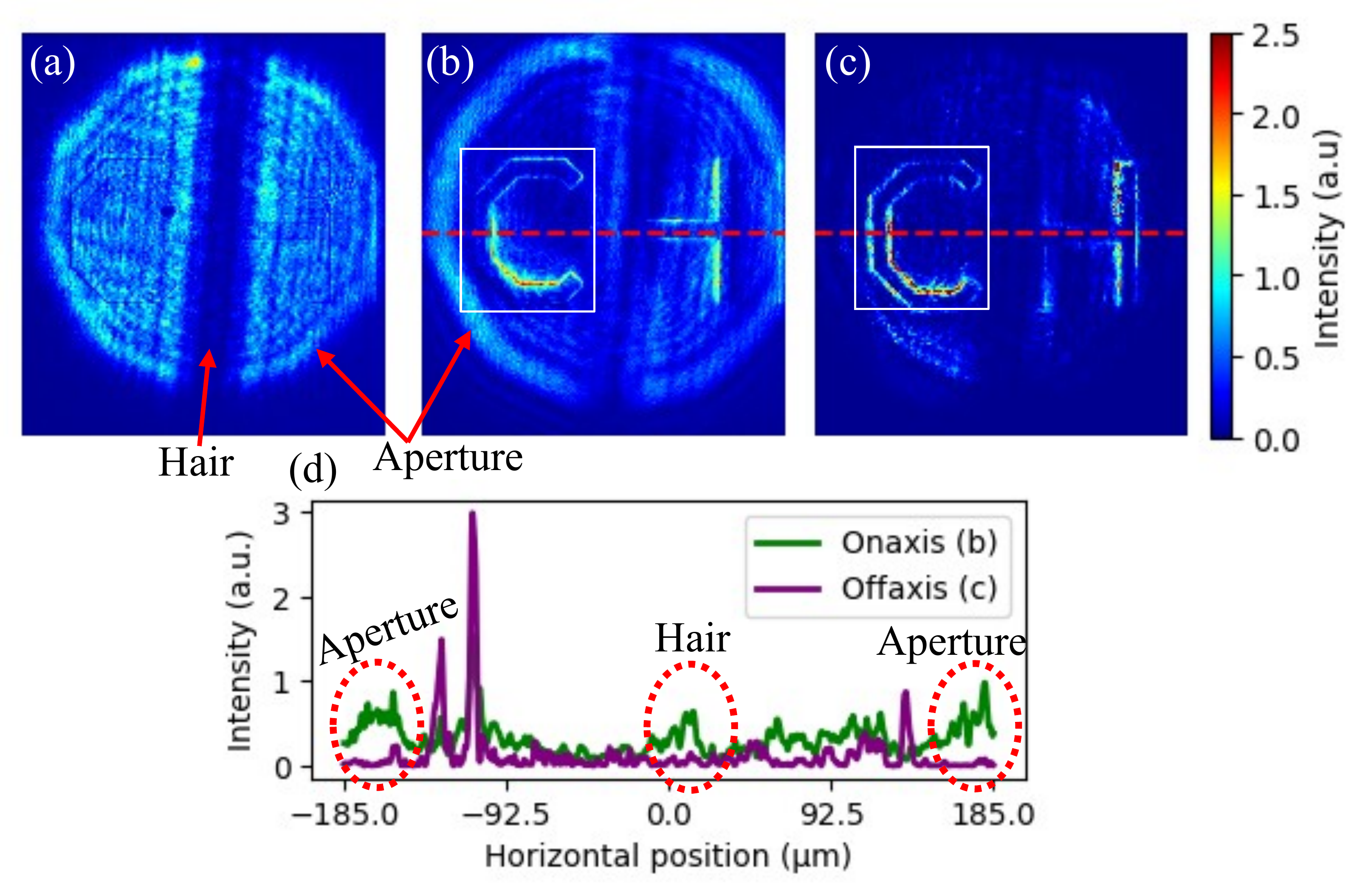}
\caption{Experimental results for the second phase-amplitude object (phase test target overlapping with a white hair and surrounded by an aperture). (a) Without q-plate, (b) on-axis edge enhancement, (c) isolated phase edge using the two-shift method, (d) horizontal cross-section along the red dotted line.}
\label{result_hair}
\end{figure}

The second phase‑amplitude object sample adds a white hair (non‑zero transmittance) that overlaps the letters “CH” of the phase test target; the aperture is also present. The experimental results are shown in Fig.~\ref{result_hair}. Without a q-plate (Fig.~\ref{result_hair}(a)), both the hair (blue strip) and the aperture are visible, but the phase test target, being transparent, is not visible. After inserting an on‑axis q-plate, the edges of the aperture, the letter “CH”, and the white hair become visible (Fig.~\ref{result_hair}(b)), though the hair edge is weak. After applying the two-shift method with $45^\circ$ and $135^\circ$ polarizers, the isolated phase edge image (Fig.~\ref{result_hair}(c)) shows a clear reduction of the hair and aperture edges, while the phase edges become more distinct. In Fig.~\ref{result_hair}(d), the horizontal cross‑section along the red dotted line in (b) and (c) is shown, with green for (b) and purple for (c). The intensity of the aperture and hair is clearly reduced in (c) compared to (b).

Table~\ref{tab:sample2} summarizes the edge reduction of the aperture and the correlation coefficient between the isolated phase edge and the on‑axis phase edge for different q-plate displacements. The correlation coefficient was calculated for the letter “C” enclosed by the white dotted rectangle.

\begin{table}[htbp]
\centering
\caption{Edge reduction of the white hair and correlation coefficient for sample 2 as a function of q-plate displacement.}
\label{tab:sample2}
\begin{tabular}{c|ccccc}
Displacement ($\mu$m) & 25 & 50 & 75 & 100 & 125 \\
\hline
Edge reduction (\%) & 87.93 & 91.66 & 92.68 & 93.88 & 93.69 \\
CC & 0.70 & 0.75 & 0.72 & 0.71 & 0.64 \\
\end{tabular}
\end{table}

The edge reduction reaches $93.9\%$ at $r_0 = 100~\mu$m, but the correlation coefficient decreases for larger displacements. This is because the on-axis reference image suffers from diffraction between the white hair and the phase test target, so the extracted phase edge used as a reference is not uniformly enhanced. However, the off-axis isolated phase edge becomes sharper and cleaner as the displacement increases. Consequently, the correlation coefficient decreases due to increasing dissimilarity with the reference. Nevertheless, the method successfully isolates the phase edges, although the phase edge overlapping with the white hair could not be fully recovered. According to Eq.~\ref{Eq5}, the isolated phase edge contains the factor $A(\bm{r})^4$ originating from the amplitude object, and as proposed in Ref.~\cite{2024isolation}, an inverse filtering deconvolution method can be used to recover the phase edge in such overlapping regions.

\section{Conclusion}
In conclusion, we successfully isolated phase edges from amplitude edges using two shifts of an off-axis q-plate with polarizers. This method is faster and simpler than the previous four-shift off-axis approach, reducing the number of measurements while maintaining effective reduction of amplitude edges. It is therefore suitable for biological imaging where distinguishing phase from amplitude objects is essential. Future work, as outlined in Ref.~\cite{2024isolation}, will apply inverse-filtering deconvolution and deep-learning methods to the output of the two-shift method to fully recover overlapping phase edges and mitigate noise.

\begin{backmatter}


\bmsection{Disclosures}
The authors declare no conflicts of interest.

\bmsection{Data availability}
Data underlying the results presented in this paper are not publicly available at this time but may be obtained from the authors upon reasonable request.

\end{backmatter}


\bibliography{ref}

\begin{thebibliography}{10}
\newcommand{\enquote}[1]{``#1''}

\bibitem{zernike1942phase}
F.~Zernike, \enquote{Phase contrast, a new method for the microscopic observation of transparent objects part {II},} {\protect\JournalTitle{Physica}} \textbf{9}, 974--986 (1942).

\bibitem{DIC1}
M.~R. Arnison, K.~G. Larkin, C.~J. Sheppard, \emph{et~al.}, \enquote{Linear phase imaging using differential interference contrast microscopy,} {\protect\JournalTitle{Journal of Microscopy}} \textbf{214}, 7--12 (2004).

\bibitem{offaxis1}
Z.~Gu, D.~Yin, S.~Nie, \emph{et~al.}, \enquote{High-contrast anisotropic edge enhancement free of shadow effect,} {\protect\JournalTitle{Applied Opt.}} \textbf{58}, G351--G357 (2019).

\bibitem{biological1}
S.~Wei, S.~Zhu, and X.~Yuan, \enquote{Image edge enhancement in optical microscopy with a {B}essel-like amplitude modulated spiral phase filter,} {\protect\JournalTitle{Journal of Optics}} \textbf{13}, 105704 (2011).

\bibitem{medical}
P.~Shankar, \enquote{Contrast enhancement and phase-sensitive boundary detection in ultrasonic speckle using bessel spatial filters,} {\protect\JournalTitle{IET Image Process.}} \textbf{3}, 41--51 (2009).

\bibitem{astonomy1}
D.~Mawet, E.~Serabyn, J.~K. Wallace, and L.~Pueyo, \enquote{Improved high-contrast imaging with on-axis telescopes using a multistage vortex coronagraph,} {\protect\JournalTitle{Optics Lett.}} \textbf{36}, 1506--1508 (2011).

\bibitem{VPM4phase1}
Y.~Liu, P.~Yu, X.~Hu, \emph{et~al.}, \enquote{Single-pixel spiral phase contrast imaging,} {\protect\JournalTitle{Opt. Lett.}} \textbf{45}, 4028--4031 (2020).

\bibitem{zangpo2026single}
J.~Zangpo, H.~Kobayashi, T.~Jinushi, and R.~Yasuhara, \enquote{Single-pixel edge enhancement of object via convolutional filtering with localized vortex phase,} {\protect\JournalTitle{Journal of Modern Optics}} \textbf{73}, 320--329 (2026).

\bibitem{2005spiral}
S.~F{\"u}rhapter, A.~Jesacher, S.~Bernet, and M.~Ritsch-Marte, \enquote{Spiral phase contrast imaging in microscopy,} {\protect\JournalTitle{Optics Express}} \textbf{13}, 689--694 (2005).

\bibitem{spiralPF4}
A.~Jesacher, S.~F{\"u}rhapter, S.~Bernet, and M.~Ritsch-Marte, \enquote{Shadow effects in spiral phase contrast microscopy,} {\protect\JournalTitle{Physical Review Letters}} \textbf{94}, 233902 (2005).

\bibitem{spiralPF7}
S.~Bernet, A.~Jesacher, S.~F{\"u}rhapter, \emph{et~al.}, \enquote{Quantitative imaging of complex samples by spiral phase contrast microscopy,} {\protect\JournalTitle{Optics Express}} \textbf{14}, 3792--3805 (2006).

\bibitem{2024isolation}
J.~Zangpo and H.~Kobayashi, \enquote{Isolation of phase edges using off-axis q-plate filters,} {\protect\JournalTitle{Optics Express}} \textbf{32}, 12911--12925 (2024).

\bibitem{qplate2}
D.~Li, S.~Feng, S.~Nie, \emph{et~al.}, \enquote{Scalar and vectorial vortex filtering based on geometric phase modulation with a q-plate,} {\protect\JournalTitle{Journal of Optics}} \textbf{21}, 065702 (2019).

\bibitem{zangpo2023edge}
J.~Zangpo, T.~Kawabe, and H.~Kobayashi, \enquote{Edge-enhanced microscopy of complex objects using scalar and vectorial vortex filtering,} {\protect\JournalTitle{Optics Express}} \textbf{31}, 38388--38399 (2023).

\bibitem{zangpoedge}
J.~Zangpo, \enquote{Edge enhancement of {P}hase-{A}mplitude {O}bject {U}tilizing {O}ptical {V}ortex in {M}icroscope,} Ph.{D}. {T}hesis (2024).

\bibitem{zangpo2022isolation}
J.~Zangpo, T.~Kawabe, and H.~Kobayashi, \enquote{Isolation of {P}hase {O}bject in {E}dge-{E}nhanced {M}icroscopy with q-plate under {T}ilted {L}aser {I}llumination,} in \emph{CLEO: Science and Innovations,}  (Optica Publishing Group, 2022), pp. JTh3B--7.

\end{thebibliography}

\end{document}